# OVERVIEW OF THE LBNE NEUTRINO BEAM*

C. D. Moore, Yun He, Patrick Hurh, James Hylen, Byron Lundberg, Mike McGee, Joel Misek, Nikolai V. Mokhov, Vaia Papadimitriou, Rob Plunkett, Ryan Schultz, Gueorgui Velev, Karlton Williams, Robert Miles Zwaska

FNAL, Batavia, IL 60510 U.S.A.

*Abstract*

The Long Baseline Neutrino Experiment (LBNE) will utilize a neutrino beamline facility located at Fermilab. The facility is designed to aim a beam of neutrinos toward a detector placed at the Deep Underground Science and Engineering Laboratory (DUSEL) in South Dakota. The neutrinos are produced in a three-step process. First, protons from the Main Injector hit a solid target and produce mesons. Then, the charged mesons are focused by a set of focusing horns into the decay pipe, towards the far detector. Finally, the mesons that enter the decay pipe decay into neutrinos. The parameters of the facility were determined by an amalgam of the physics goals, the Monte Carlo modeling of the facility, and the experience gained by operating the NuMI facility at Fermilab. The initial beam power is expected to be ~700 kW, however some of the parameters were chosen to be able to deal with a beam power of 2.3 MW.

## INTRODUCTION

The LBNE neutrino beam needs to provide a wide band beam to cover the first and second neutrino oscillation maxima (.5 GeV to 4 GeV) and modeling has been used in the choice of system parameters to optimize the operation of the facility in this region. The initial operation of the system will be at a beam power incident on the production target of 700 kW however some of the initial implementation will have to be done in such a manner that operation at 2.3 MW can be achieved without retrofitting. The cooling systems for the chase, decay tunnel, and absorber must be designed so that an upgrade for 2.3 MW operations is achievable without cost implications other than increasing the capacity. The relevant radiological concerns, prompt dose, residual dose, air activation, tritium production have been extensively modeled and these issues have been used in the system design. This paper is a snapshot of the present status of the design as exemplified in the current Conceptual Design Report [1].

## MODELING

A software model for the entire beamline has been coded into the MARS15 simulation package [2]. This model includes all the essential components of the target chase, decay channel, hadron absorber and the steel and concrete shielding in the present design. The MARS simulations are used specifically in the calculation of: (1) beam-induced energy deposition in components for engineering design, (2) prompt dose rates in halls and outside shielding, (3) residual dose rates from activated components, (4) radionuclide production in components, shield and rock, (5) horn focusing design and optimization of neutrino flux. Fig. 1 shows the MARS model of the target chase. Calculated dynamic heat loads in the hottest components are 14 to 30 kW, with residual dose rates on contact ranging from 145 to 345 Rem/hr after 30-day irradiation and 1-day cooling.

A most important aspect of modeling at the present design stage (Conceptual Design level 1 or "CD-1") is the determination of necessary shield thickness and composition. For example, ground water protection is a crucial consideration for the 250 m-long decay tunnel. Results from modeling calculations are essential in the design of all of the subsystems discussed in the following sections.

The neutrino beam is a set of components and enclosures designed to efficiently convert the primary proton beam (up to 120 GeV) to a neutrino beam (maximum flux at neutrino energies 2-3 GeV) aimed at the far detectors, 1280 km away. In order of placement, this design includes (1) a target and horn protection system, (2) toroidal focusing horns for the secondary pions and surrounding shielding, (3) a 250m long, 4m diameter decay pipe to allow pions to decay to neutrinos, and a (5) absorber at the end of the decay pipe to deal with uninteracted protons and residual pions. The level of detail incorporated into the present simulations is conveyed in Fig. 1.

___________________________________________
*Work supported by the Fermilab Research Alliance, under contract DE-AC02-07CH11359 with the U.S. Dept of Energy.
[dagger]cmoore@fnal.gov

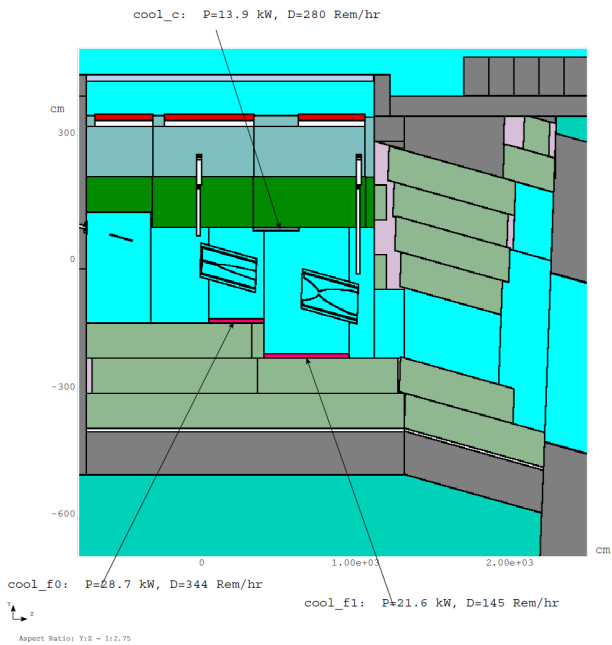

Figure 1. MARS model of the target chase with horns, shielding and upstream end of the decay channel.

## BEAM WINDOW, BAFFLE, AND TARGET DESIGN

The conceptual primary beam window design considers a 50 mm diameter, 0.2 mm thick partial hemispherical beryllium window water-cooled given the 2.3 MW (1.5 mm spot size) case. A stainless steel, 117.5 mm O.D. Conflat flange with knife-edge seal provides a bolted connection to the primary beam pipe.

The baseline baffle design considers 700 kW beam energy with a view towards the beam power upgrade to 2.3 MW. The baseline baffle consists of ten 57 mm O.D. x 13 mm I.D. x 150 mm long graphite R7650 grade cores which are enclosed by a 150 cm long aluminum tube after annealing. The graphite baffle prevents mis-steered primary proton beam from causing damage to the horn neck and target cooling/support components. It must withstand the full intensity of the beam for a few pulses during the time needed to detect the mis-steered beam and terminate beam.

The conceptual target design for LBNE is based on experience with the Fermilab NuMI neutrino beam and studies for higher power beams. Graphite was used in the NuMI beam, and its performance was basically successful. However, over time four targets have failed due to ancillary components, and one has failed from ~ 10% graphite degradation. Graphite has been adopted as the baseline target material, but alternatives are under study.

The target uses a graphite core evolved from the NuMI designs and studies for a higher-power beam, developed at IHEP-Protvnio. The graphite core is segmented into short cylinders 15.3 mm in diameter and 25 mm in length. The total graphite length is 95 cm. The cylinders are shrink-fit within a stainless steel tube. The tube applies a pre-stress to the graphite, which has stronger compressive than tensile strength. The encapsulated core is surrounded by a double-annulus cooling tube, through which water flows to remove the heat.

## HORNS AND HORN POWER SUPPLY

A half-sine wave 300 kA pulsed power supply drives the two series connected horns via direct stripline coupling from a capacitor bank. Capable of cycling at 1.33 Hz continuously, features include electronic control of horn current polarity as rapidly as pulse to pulse, energy recovery, plus many proven design aspects of the NuMI horn system currently operating at Fermilab.

The focusing of charged pions is produced by the toroidal magnetic field present in the volume between the co-axial inner and outer conductors of the horns with a pulsed current. The inner conductor of Horn 1 has a straight cylindrical section upstream surrounding the target, followed by a parabolic section downstream. Horn 2 is a double-paraboloid.

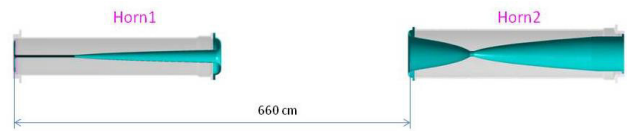

Figure 2. LBNE Horns

Horn conductors must withstand repetitive thermal and magnetic loading over millions of beam/current pulses. The conductors are made of aluminum 6061-T6 and will be cooled by spray water. Thermal and structural finite element analyses were carried out to guide the design and study the fatigue strength of the inner conductors. Horns will be supported and positioned by support modules, which are capable of aligning and servicing the horn by remote control.

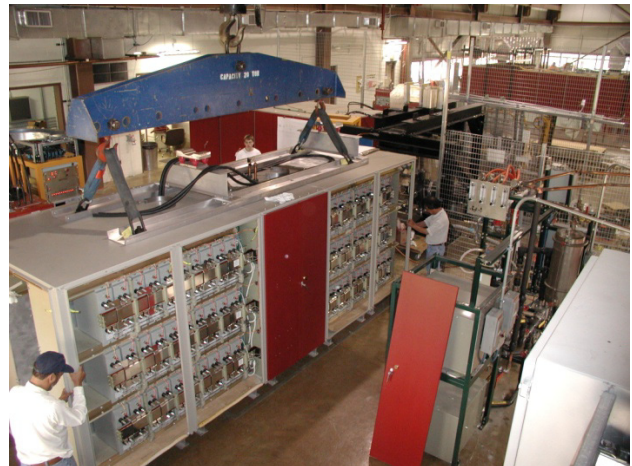

Figure 3. Completed NuMI horn power supply.

## TARGET PILE SHIELD PILE

The shield surrounds the baffle, target and horns and leads directly to the decay pipe. The shield's purpose is to provide radiological protection for groundwater concerns, electronic component lifetime issues, and personal access issues with respect to residual radiation. It is composed of an inner layer of steel surrounded by concrete.

## DECAY PIPE AND HADRON ABSORBER

The decay pipe is the region where the mesons generated from the target decay into neutrinos. The dimensions (4m diameter and 250m length) of the pipe have been chosen based upon considerations of cost and maximizing of the neutrino flux in the desired energy range.

As by-product of neutrino production, a flux of primary protons (~ 15% of protons) and non-decayed secondary hadrons (mostly $\pi$ and K -mesons) and leptons must be absorbed to prevent them from entering the surrounding rock of the excavation and inducing radioactivity. This is accomplished with a specially absorber structure which is located directly after the ~ 250 m long decay pipe (DP). It is a pile of aluminum (Al), steel and concrete blocks, some of them water-cooled, which must contain the energy of the particles after the DP. The vast majority of these secondary hadrons and primary protons, essentially, all of them are stopped in the Absorber. The fluxes of secondary particles (mainly neutrons) escaping the system must be attenuated by the Absorber and shielding to the tolerable levels. According to the simulation, in total, 21% of the total beam power is deposited in the Absorber where the primary protons deposit approximately 80% of this energy. In details, the absorber design is described in [3].

## RAW WATER SYSTEMS

Radioactive Water (RAW) systems are small volume systems that are used to take heat away from regions where much thermal heat load is developed. The systems must be robust and have containment capability for the entire capacity of the system.

General construction will be to ANSI Code B31.1 Process Piping, with rigorous weld inspection and radiography. Reservoir tanks will be specified as coded pressure vessels. Redundancy and containment will be used where deemed appropriate. Significant filtration and deionizing bottles will be used to minimize the particulate build-up in the water. RAW capture and makeup systems will be integrated into the design, so that down time and worker exposure for routine service may be minimized.

Intermediate systems will transfer heat from the RAW systems, to chilled water systems that take the heat to the surface for removal via chillers and/or cooling ponds.

## REMOTE HANDLING EQUIPMENT

Technical components installed in the Target Chase and in the Absorber Hall areas are subjected to intense radiation from the primary or secondary beam. The level of irradiation in some LBNE environments will reach levels that are unprecedented at Fermilab. Components to be handled, serviced, and/or stored range in size (from 0.25 $m^3$ to 25 $m^3$), weight (from 10 kg to 30,000 kg), and estimated dose rate (from 5 R/hr to 8000 R/hr on contact). Therefore remotely operated removal and handling systems are an integral part of the Target Hall design. Since the remote handling systems are integrated into the infrastructure of the Target Hall and cannot be upgraded after irradiating the Target Hall and Absorber Hall areas, they must be designed to be sufficient for 2+ MW beam power.

Remote Handling Systems will include an in-situ, remotely operated target replacement rig and a work cell enabling horn replacement (with tele-manipulator stations) in the underground Target Hall area. In addition, a near surface facility will include a remote handling procedures mock-up area, a maintenance work cell (to allow component repair or volume reduction for storage) and a long-term storage area with capacity for 24 radioactive components. Cranes in the remote handling service areas are expected to include redundant drives to mitigate hazards associated with crane drive failure while moving a radioactive component.

## TRITIUM MITIGATION

Based upon experience with NuMI a system has been designed that will mitigate the expected production of tritium in such a manner that no detectable amount of tritium will be in the waters leaving the site boundaries.

## √ SUMMARY

A proof of principle design has been pushed through to provide an example of a system that would provide a desired neutrino flux at the DUSEL location. Work is in progress to optimize the costs while still keeping a viable physics program.